\documentclass[%
 reprint,
 amsmath,amssymb,
 aps,
]{revtex4-2}

\usepackage{graphicx}
\usepackage{dcolumn}
\usepackage{bm}
\usepackage{siunitx}
\usepackage{tabularx}
\usepackage{appendix}
\usepackage{amsmath,array}
\usepackage{kantlipsum}
\usepackage{float}
\usepackage{multirow}
\usepackage{adjustbox}
\usepackage{booktabs}
\usepackage{stackengine}
\usepackage[T1]{fontenc}

\begin{document}

\preprint{APS/123-QED}

\title{Low-loss GHz frequency phononic integrated circuits in Gallium Nitride for compact radio-frequency acoustic wave devices}
\author{Mahmut Bicer}
\email{mahmut.bicer@bristol.ac.uk}
\address{Quantum Engineering Technology Labs and Department of Electrical and Electronic Engineering, University of Bristol,
Woodland Road, Bristol BS8 1UB, United Kingdom}

\author{Krishna C. Balram}
\email{krishna.coimbatorebalram@bristol.ac.uk}
\address{Quantum Engineering Technology Labs and Department of Electrical and Electronic Engineering, University of Bristol,
Woodland Road, Bristol BS8 1UB, United Kingdom}

\date{\today}
\begin{abstract}
Guiding and manipulating GHz frequency acoustic waves in \si{\um}-scale waveguides and resonators opens up new degrees of freedom to manipulate radio frequency (RF) signals in chip-scale platforms. A critical requirement for enabling high-performance devices is the demonstration of low acoustic dissipation in these highly confined geometries. In this work, we show that gallium nitride (GaN) on silicon carbide (SiC) supports low-loss acoustics by demonstrating acoustic microring resonators with frequency-quality factor ($fQ$) products approaching \qty{4e{13}}{\Hz} at 3.4 GHz. The low dissipation measured exceeds the $fQ$ bound set by the simplified isotropic Akhiezer material damping limit of GaN. We use this low-loss acoustics platform to demonstrate spiral delay lines with on-chip RF delays exceeding \qty{2.5}{\us}, corresponding to an equivalent electromagnetic delay of $\approx$ \qty{750}{\m}. Given GaN is a well-established semiconductor with high electron mobility, our work opens up the prospect of engineering traveling wave acoustoelectric interactions in \si{\um}-scale waveguide geometries, with associated implications for chip-scale RF signal processing.
\end{abstract}

\maketitle

\section{I\MakeLowercase{ntroduction}}\label{sec1}

Gigahertz (\unit{\GHz}) frequency acoustic wave devices underpin a variety of critical classical and quantum information processing technologies, ranging from RF filters in modern smartphones \cite{lam2016review} to the development of quantum transducers for interconnecting remote superconducting qubit-based processors \cite{balram2022piezoelectric, bardin2021microwaves}. Building on developments in (silicon) photonic integrated circuits on low-loss on-chip routing and manipulation of light at telecommunication wavelengths ($\lambda_{o}\approx$ \qty{1550}{\nm}) in sub-\si{\um} waveguides and resonators \cite{chrostowski2015silicon},  there has been a general trend towards extending the same degree of control towards high-frequency acoustic waves \cite{safavi2019controlling}, given the similarity in wavelength ($\approx$ \qty{1}{\um}) between the two fields. In a piezoelectric material, the acoustic wave is usually generated from an RF signal of interest, and therefore, manipulating acoustic waves in chip-scale platforms provides new degrees of freedom for RF signal manipulation, ranging from exploiting non-reciprocity in RF signal processing \cite{nagulu2020non} to the development of integrated front-ends that can support full duplex wireless protocols \cite{sabharwal2014band}.  

To realize their full potential, especially from an RF systems perspective, these GHz phononic integrated circuits (PnICs) platforms \cite{fu2019phononic, mayor2021gigahertz, bicer2022gallium}  need to achieve end-to-end insertion losses that are comparable to traditional bulk and surface acoustic wave devices \cite{hashimoto2009rf, datta1986surface, morgan2010surface, bhugra2017piezoelectric}. This requires addressing two key challenges: (i) showing that GHz acoustic waves can be focused into and out of \si{\um}-scale waveguides with near unity efficiency \cite{siddiqui2018lamb, bicer2022gallium}, and (ii) demonstrating that acoustic waves can be routed in these \si{\um}-scale waveguide geometries with negligible excess dissipation and scattering. 

In this work, we focus on the second problem and demonstrate that high-frequency ($\text{>}$ \qty{ 3}{\GHz}) acoustic waves can be guided with negligible excess dissipation and scattering in GaN PnICs. We use GaN-on-SiC \cite{bicer2022gallium} as our platform of choice for these experiments. GaN, in contrast to traditional acoustic wave devices \cite{bhugra2017piezoelectric} which exploit stronger piezoelectric insulators like scandium aluminum nitride (ScAlN) and lithium niobate (LN), is a moderate piezoelectric semiconductor with high electron mobility \cite{rais2014gallium}. This combination opens up the possibility of engineering acoustoelectric interactions in \si{\um}-scale waveguide geometries \cite{ghosh2019acoustoelectric, hackett2021towards}, enabling tight integration between passive acoustic components and active semiconductor devices on the same die. We would like to note here that while the idea of guiding acoustic waves in waveguides is not new \cite{oliner1976waveguides}, and 1D wave confinement has been exploited in commercial devices \cite{takai2017ihp, hagelauer2022microwave}, our main focus here is on waveguide geometries with strong transverse confinement with waveguide dimensions $\sim \lambda^2_a$, where $\lambda_a$ is the acoustic wavelength, analogous to silicon integrated photonics.   

We show that GaN-on-SiC acoustic microring resonators can achieve figures of merit, represented by the $fQ$ product of the resonator \cite{bhugra2017piezoelectric}, of   \qty{ 4e{13}}{\Hz} at \qty{3.4}{\GHz} operating frequency at ambient conditions in our best devices. This $fQ$ product exceeds the isotropic Akhiezer damping limit (AKE) of GaN \cite{tabrizian2009effect, ghaffari2013quantum}, which provides a rough upper bound on the achievable quality factor at a given operating frequency. By relying on whispering gallery modes, which use the total internal reflection of sound for acoustic confinement, it is possible to minimize the excess acoustic dissipation and scattering. Furthermore, the fundamental material limited quality factors can be achieved while maintaining all the advantages of tight (\si{\um}-scale) confinement and manipulation in a guided wave geometry. We harness this low on-chip propagation loss to demonstrate integrated spiral delay lines that achieve RF signal delays of \qty{2.5}{\us}, corresponding to an equivalent electromagnetic delay of $\approx$\qty{750}{\m}. Provided the insertion loss problem (i, above) can be similarly addressed through the design of efficient \si{\um}-scale focusing transducers \cite{siddiqui2018lamb, bicer2022gallium}, our work opens up the possibility of re-thinking RF systems from the ground up around PnICs.

\section{M\MakeLowercase{icroring Resonators with }$fQ$ \MakeLowercase{products approaching the \MakeUppercase{A}khiezer damping limit}}\label{sec2}

\begin{figure}[!hbtp]
{\includegraphics[width = 1.0 \columnwidth]{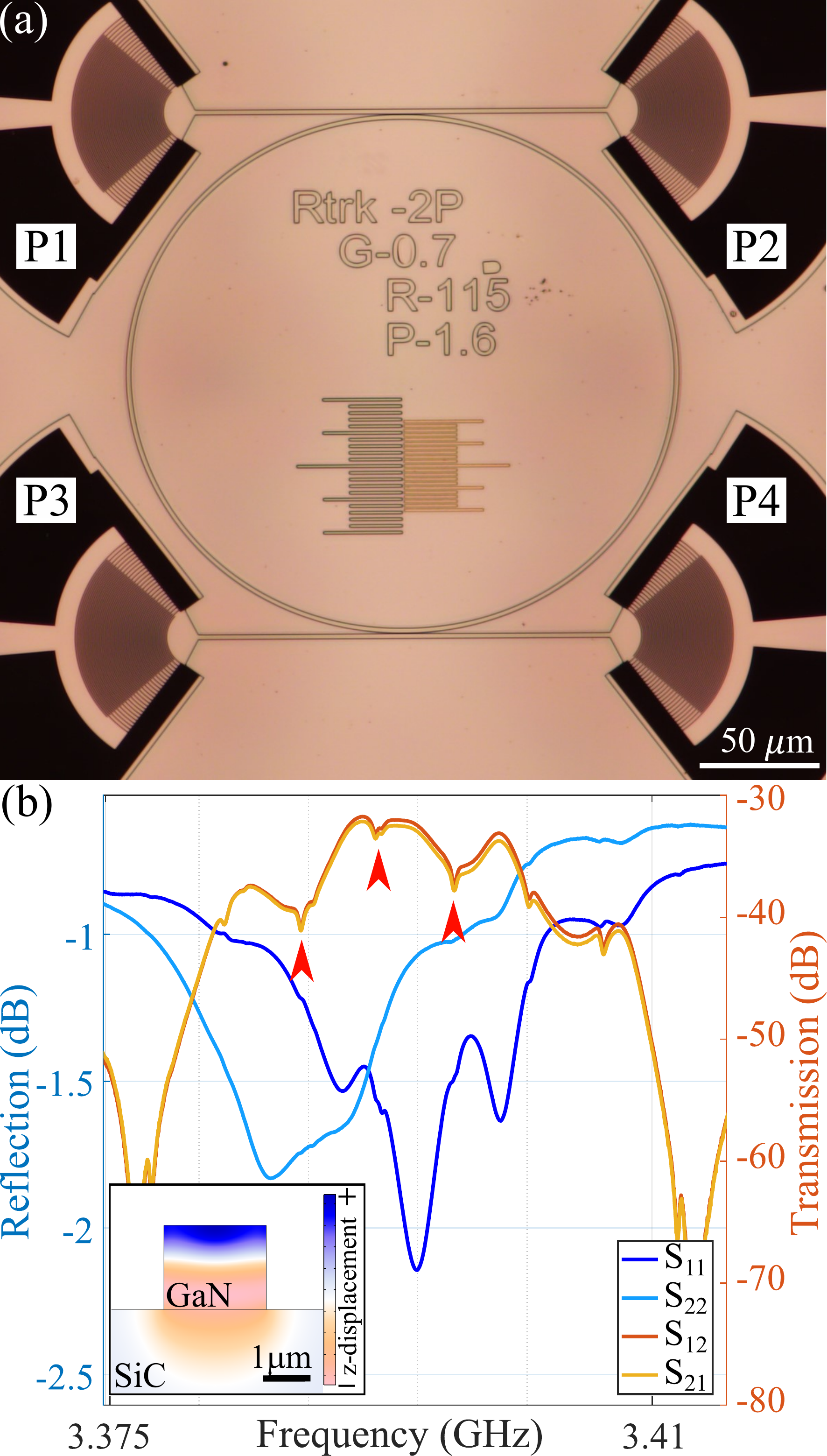}}
  \caption{(a) Optical image of 4-port ring device, with a ring radius ($R$) of 115 \si{\um}, 700 \si{\nm} ($g_{wvg}$) spacing between waveguide and ring, and coupling length ($L_{cp}$) of $2\lambda_a$. IDTs' period ($\lambda_a$) is 1.6 \si{\um} while waveguides' and ring's width ($w_{wvg}$) are 1.8 \si{\um} and GaN thickness ($t_{wvg}$) is 1.5 \si{\um}, (b) Frequency spectrum of a 4-port device with ring $R = 115$ \si{\um}, an $L_{cp}$ = $2\lambda_a$, and $g_{wvg} = 500$ \si{\nm}. The transmission ($S_{21}$ and $S_{12}$) and reflection ($S_{11}$ and $S_{22}$) of ports P1 and P2 are presented in the log-scale. The red arrows highlight the ring dips and the inset presents the Lamb wave mode profile.}
  \label{fig:1}
\end{figure}

Figure \ref{fig:1}(a) shows an optical microscope image of a fabricated four-port symmetric acoustic microring resonator. \qty{3.4}{\GHz} Lamb waves are launched into acoustic waveguides using focusing interdigitated transducers (IDT, P1-4) \cite{bicer2022gallium}. The acoustic waveguides are side-coupled to the microring resonator, and nominally identical focusing IDTs are fabricated to act as acoustic receivers in the through, cross, and drop ports (labeled P1, P2, P3, and P4 in Fig.\ref{fig:1}(a)).  The device is fabricated on a GaN-on-SiC wafer sourced from IQE with a thin (\qty{< 50}{\nm}) aluminum nitride (AlN) buffer layer. The GaN layer is iron doped to make it insulating. The fabrication process is similar to our previous work \cite{bicer2022gallium}, and employs a multi-layer aligned electron beam lithography process to define the waveguides and IDT. The reactive ion etching and cleaning steps have been optimized to minimize surface roughness and sidewall residue in order to reduce the overall scattering loss in the waveguides. 

The IDT electrode period ($\lambda_a$), the GaN layer thickness ($t_{wvg}$), and the waveguide (and ring resonator) width ($w_{wvg}$) are kept constant across all devices at 1.6 \si{\um}, 1.5 \si{\um}, and 1.8 \si{\um}, respectively. With a view towards achieving critical coupling \cite{yariv2002critical} between the waveguide and the resonator, the waveguide resonator gap ($g_{wvg}$) is varied from 500 \si{\nm} to 800 \si{\nm} in 100 \si{\nm} steps for a fixed coupling length ($L_{cp}$) of $2\lambda_a$, and the $L_{cp}$ is varied from $2\lambda_a$ to $10\lambda_a$, in $4\lambda_a$ steps for a constant $g_{wvg}$ of 500 \si{\nm}. Figure \ref{fig:1}(b) shows the measured through-port transmission spectrum ($S_{21}$) of a representative device with a radius ($R$) of 115 \si{\um}, $L_{cp}$ = $2\lambda_a$, and $g_{wvg}$ = 500 \si{\nm}. The transmitter and receiver IDT characteristics are also shown in the overlaid RF reflection ($S_{11}$) spectra. While the IDTs are nominally designed to be identical, the measured characteristics are different due to fabrication imperfections, mainly substrate charging and proximity error correction effects. The main features of interest are the sharp dips (indicated by the red arrow in Fig.\ref{fig:1}(b)) that correspond to successive acoustic whispering gallery resonances of the microring resonator. 

\begin{figure}[!hbtp]
{\includegraphics[width = 1.0 \columnwidth]{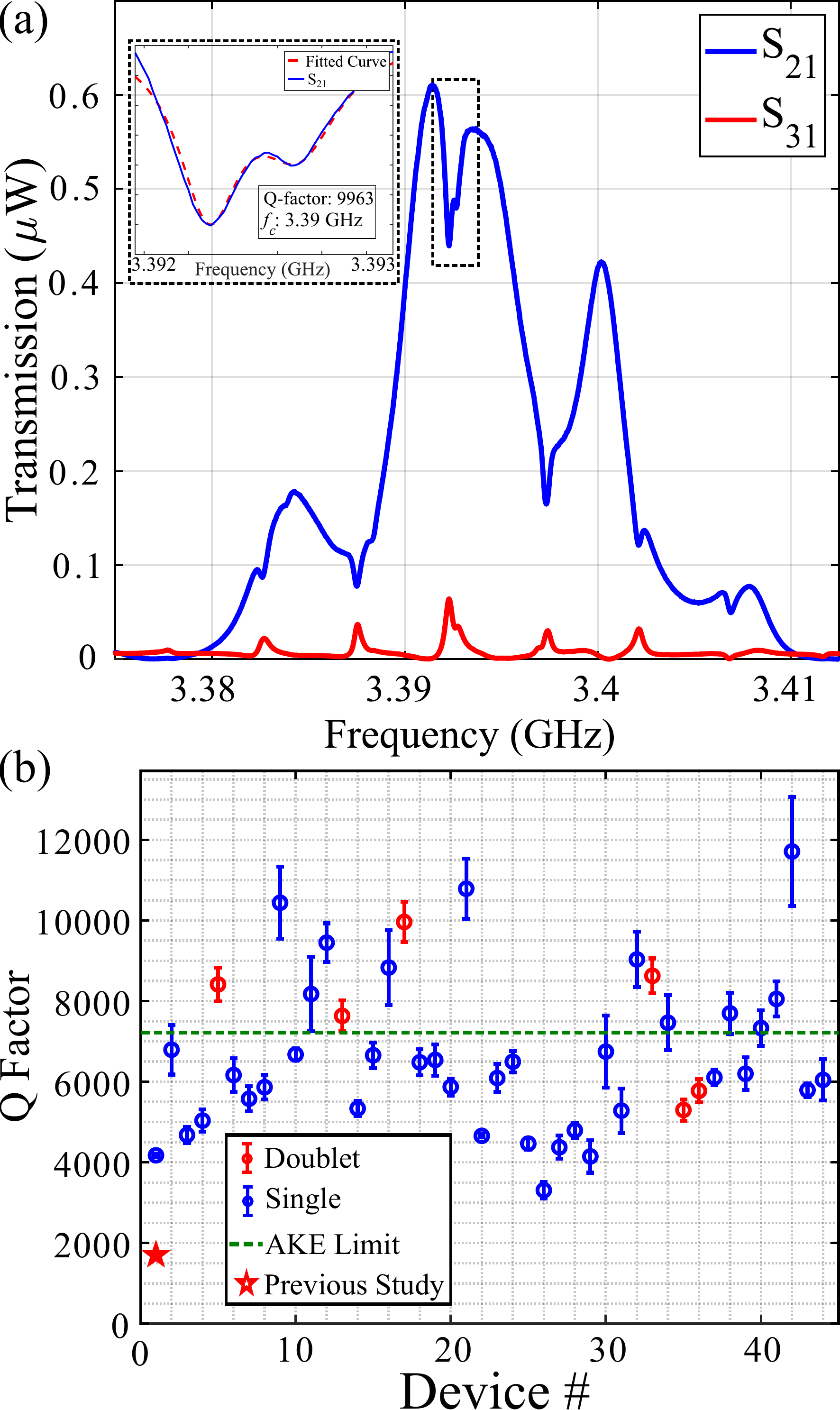}}
  \caption{(a) Transmission spectrum of a 4-port device with ring $R = 115$ \si{\um}, an $L_{cp}$ = $2\lambda_a$, and $g_{wvg} = 500$ \si{\nm}. The transmission of through (blue) ($S_{21}$) and drop (red) ($S_{31}$) are presented in linear scale. The Q-fitting of the doublet ring dips is presented in the inset. The red dash curve presents the fitting while the solid blue line presents the data highlighted with the dashed box, (b) $Q$ factors of our fabricated devices with the fitting error indicated with bars. The green line indicates the GaN Akhiezer $fQ$ limit for a frequency of \qty{3.4}{\GHz}.}
  \label{fig:2}
\end{figure}

\begin{figure}[!hbtp]
{\includegraphics[width = 1.0 \linewidth]{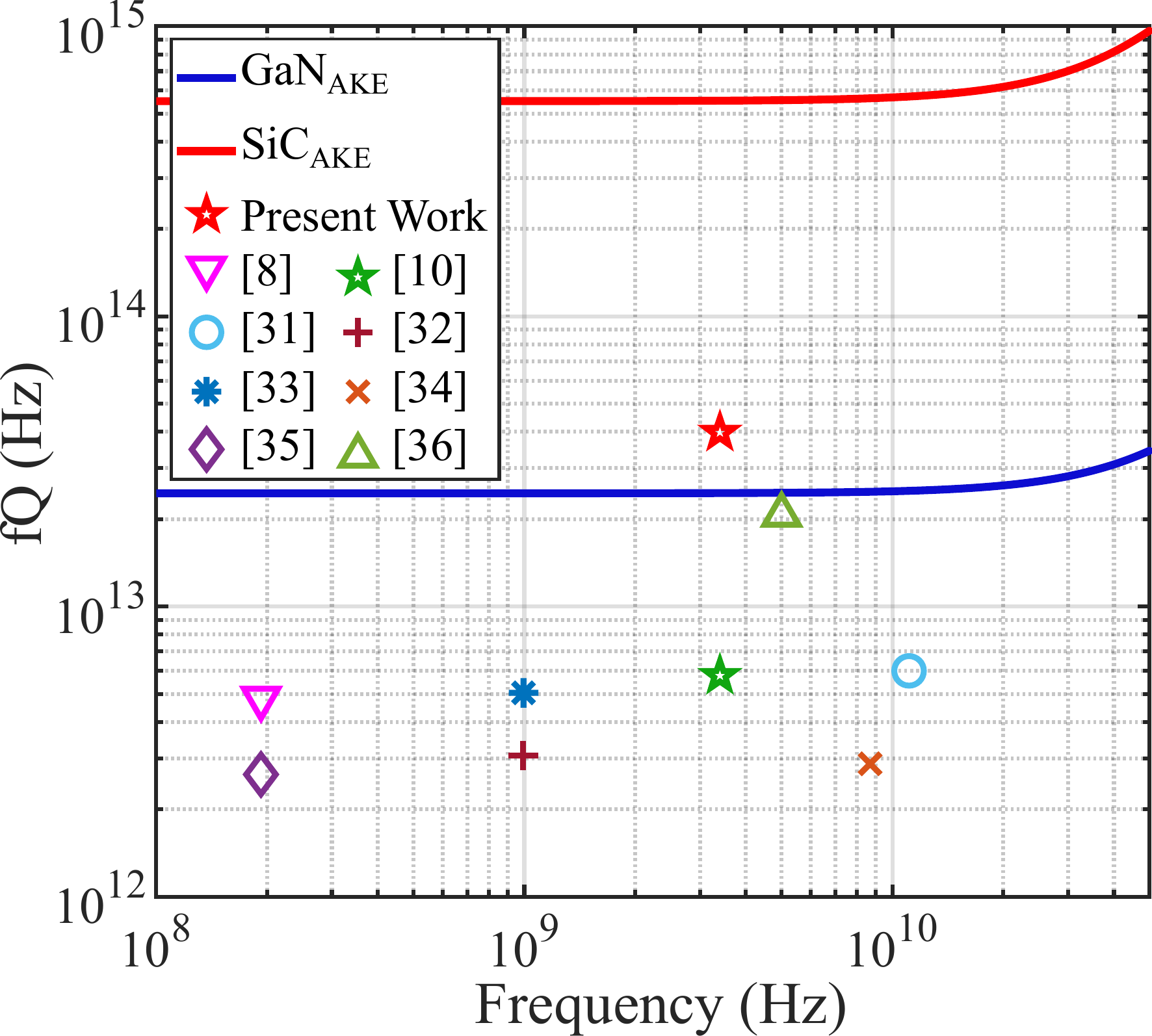}}
  \caption{ Akhiezer $fQ$ limit of SiC and GaN materials indicated in red line and blue line, respectively, and some of the reported GaN $fQ$ values in the literature \cite{bicer2022gallium,fu2019phononic,ahmed2023super,wang2014gan,ahmed2021switchable,ansari20148,xu2018high,xu2022high}. The references are placed in the main text. The maximum $fQ$ product device is reported in this study and it is indicated with the red star (\qty{3.98e{13}}{\Hz}) while the $fQ$ product of the previous study is indicated with the green star.}
  \label{fig:3}
\end{figure}

The acoustic transmission spectrum of the device in both the through and drop ports is plotted again (in linear scale), in Fig.\ref{fig:2}(a), showing good correspondence between the frequencies of the dips in the through-port and the peaks in the drop-port, as expected. The inset in Fig.\ref{fig:2}(a) shows a zoomed-in spectrum of one of the modes of interest, clearly showing a doublet. In keeping with the optics analogy \cite{weiss1995splitting}, the appearance of doublets in the transmission spectrum is usually an indicator of the presence of high-quality factor ($Q$) modes in the system. We estimate the $Q$ of the modes using a Lorentzian lineshape fit \cite{petersan1998measurement} to the (power) spectrum and estimate a loaded $Q$ of 9,963 with a 5\% fitting error. We find high $Q$ modes across a wide variety of devices, as indicated in Fig.\ref{fig:2}(b). The data points reported in our previous work \cite{bicer2022gallium} are also indicated to show the improvement in $Q$, obtained through fabrication optimization. The highest $Q$ measured in these devices at ambient conditions is 11,710. This corresponds to an $fQ$ product of \qty{3.98e13}{\Hz}, which is by far the highest reported in the GaN material platform as shown in Fig. \ref{fig:3}. Such high $Q$'s point to the importance of designing acoustic resonators around whispering gallery modes with the total internal reflection of sound, rather than relying on metallic boundaries or grating reflectors, which always provide excess dissipation and scattering, especially at GHz frequencies.

To provide some context on the ultimate dissipation limits and the prospects for future improvement, it is instructive to compare the $Q$ factors achieved in these experiments with the $fQ$ bounds set by AKE in GaN \cite{ghaffari2013quantum, tabrizian2009effect}. We use a simplified expression for the AKE limit that is applicable to isotropic media at a given frequency ($\omega=2{\pi}f$):

\begin{eqnarray}
\label{eqn:1}
    fQ = \frac{\rho v_{a}^2 (1+(\omega\tau)^2) }{2 \pi \gamma^2 C_v T \tau} 
\end{eqnarray}

where the material parameters and their corresponding values for GaN and SiC \cite{rais2014gallium}) are presented in Table \ref{table:1}.
\begin{table*}
    \centering
    \begin{tabular}{||c|c|c|c|c||} 
     \hline
     Material Property & GaN & SiC & Units & Ref. \\ 
     \hline\hline
     Mass density  ($\rho$) & \num{6150} & \num{3210} & \unit{kg/m^3} & \cite{siklitsky1998semiconductors} \\ 
     Specific heat  ($C_s$) & \num{490} & \num{750} & \unit{J/kg.K} & \cite{siklitsky1998semiconductors} \\
     Electron mobility (bulk)  ($\mu$) & \num{200} & \num{400} & \unit{cm^2/V.s} & \cite{siklitsky1998semiconductors} \\
     Free electron concentration ($N$) & \num{5e{15}} & $-$ & \unit{cm^{-3}} & \cite{popa2015gallium} \\
     Coupling coefficient  ($\textbf{k}^2$) & 0.02 & \num{8e{-4}} & $-$ & \cite{rais2014gallium} \\ 
     Dielectric constant ($\epsilon_r$) & \num{9.5} & \num{9.66} & $-$ & \cite{siklitsky1998semiconductors} \\
     Phonon relaxation time ($\tau$) & \num{2e{-12}} & $-$ & \unit{\second} & \cite{rais2014gallium} \\
     Thermal conductivity ($\kappa$) & \num{130} & \num{360} & \unit{W/m-K} & \cite{siklitsky1998semiconductors} \\
     Linear expansion coefficient   ($\chi$) & \num{5.6e{-6}} & \num{2.77e{-6}} & \unit{K^{-1}} & \cite{siklitsky1998semiconductors} \\
     Acoustic velocity ($v_a$) & \num{7960} & \num{13300}& \unit{m/s} & \cite{siklitsky1998semiconductors,tabrizian2009effect} \\
     Grüneisen parameter ($\gamma$) & \num{1.18} & \num{0.3} & $-$ & \cite{perlin1992raman,ghaffari2013quantum} \\
     \hline
     \end{tabular}\\
    \caption{Wurtzite GaN and SiC material properties used for AKE limit calculation in equation \ref{eqn:1}}
    \label{table:1}
    \end{table*}
    
The use of the AKE limit ($\omega\tau<<1$) to bound the dissipation is appropriate for the device frequencies ($<$ 4 GHz) considered in this work because the relevant phonon scattering time (${\tau}$) for GaN is estimated to be \qty{2.02}{\ps} \cite{rais2014gallium}. Figure \ref{fig:3} plots the measured $Q$ of our device, along with a few other GHz frequency acoustic resonators in GaN from the literature. We overlay the estimated AKE limit for both GaN and SiC using equation \ref{eqn:1}. As can be seen, the devices reported in this work show the highest $fQ$ products by far for any reported GaN devices, and the best-performing devices exceed the AKE limit of GaN, as expressed by equation \ref{eqn:1}. We would like to note here that the AKE limit as expressed by equation \ref{eqn:1} should be interpreted more as a rough estimate for the dissipation and not a hard bound on the achievable quality factors as some of the assumptions underlying equation \ref{eqn:1}, in particular, isotropic material parameters and mode independent dissipation \cite{iyer2016mode}, are not true of the devices here.

While equation \ref{eqn:1} is a simplification, an interesting question to ponder is why the $Q$ factors of the devices in this work even approach the AKE limit in the first place, given that the interface between GaN and SiC is known to have threading dislocations due to lattice mismatch. We believe this is currently due to a combination of two factors: (a) the size (density contrast) of the threading dislocations is deeply sub-wavelength ($<$ 8 \si{\nm} compared to the acoustic $\lambda_a$ of 1.6 \si{\um}), and their density is very low ($\approx10^8$ \si{\cm^{-2}} $\sim$ \qty{1}{\um^{-2}}) in these high-quality GaN-on-SiC substrates, which results in minimal acoustic wave-dislocation interaction and scattering, and (b) a fraction of the mode's acoustic energy (mode shape shown in Fig.\ref{fig:1}(b) inset) resides in SiC which has one of the highest $fQ$ products amongst known materials \cite{ghaffari2013quantum}. We believe the high $fQ$ products measured here more accurately reflect a weighted average between the GaN and SiC values as indicated in the inset in Fig.\ref{fig:1}(b), although this needs to be confirmed experimentally.

In addition to high $Q$, the PnIC micro-rings, from an RF systems perspective, it is equally important to operate these microrings near critical coupling, wherein the coupling rate from the waveguide to the resonator equals the intrinsic decay rate as the sound propagates in the ring \cite{yariv2002critical}. For the fabricated devices reported in this work, we are currently operating in an under-coupled regime. Using temporal coupled mode theory \cite{jd2008photonic}, we can relate the measured device $Q$ in the drop port ($S_{31}$) to the resonator's intrinsic $Q$ ($Q_i$), the waveguide $Q$ ($Q_w$) that captures the resonator waveguide coupling, and the overall signal transmission magnitude, see Appendix \ref{secA1} for derivation. 

\begin{equation}\label{eqn2}
    S_{31}(\omega_0) = \frac{P_{drop}}{P_{in}} = \left ( \frac{Q}{Q_w} \right)^2 =  \left(\frac{Q_i}{Q_w+Q_i}\right)^2
\end{equation}\\ 

Equation \ref{eqn2} enables us to estimate $Q_i$ and $Q_w$ by extracting $Q$ and $S_{31}$ from the experimental data:

\begin{equation}
    Q_w = \sqrt{\frac{Q^2}{S_{31}(\omega_0)}}
\end{equation}
\begin{equation}
    Q_i = \frac{QQ_w}{Q_w - Q}
\end{equation}\\

\par The fitted $Q$s and calculated $Q_i$ and $Q_w$ of the microring with three different coupling lengths ($L_{cp}$) are given in Table \ref{table:2}. From the table, we can see $Q_w >> Q_i$ which means the microrings are heavily under-coupled for our current waveguide resonator gaps ($g_{wvg}$) of \qty{500}{\nm}. The reduction in $Q_w$ with increasing $L_{cp}$ also verifies this. Based on FEM simulations, we estimate $g_{wvg}\approx$ \qty{100}{\nm} and $L_{cp}\approx$ $71\lambda_a$ to achieve critical coupling in these devices. The lack of an evanescent acoustic field in the air makes critical coupling more challenging to achieve in PnICs than their photonic counterparts, although the geometries are well within reach of modern nanofabrication methods.

    \begin{table}[!hbtp]
    \centering
    \begin{tabular}{|| c | c | c | c | c ||} 
     \hline
     Fitted $Q $ & $Q_i$ & $Q_w$ & $L_{cp}$ & $\alpha$ \\ 
     \hline\hline
     
     $9963$ & 10043 & $124,538$ & $2\lambda_a$ & \qty{2.4}{\dB\per\mm}\\ 
     $6446$ & 6495 & $846,340$ & $2\lambda_a$ & \qty{3.7}{\dB\per\mm} \\ 
     $6256$ & 6314 & $655,592$ & $6\lambda_a$ & \qty{3.8}{\dB\per\mm}\\ 
     $4455$ & 4505 & $398,440$ & $10\lambda_a$ & \qty{5.4}{\dB\per\mm}\\ 
     \hline
    \end{tabular}
    \caption{Calculated $Q$, $Q_i$, $Q_w$, and loss parameter ($\alpha$), for 3 different coupling lengths of $2\lambda_a$, $6\lambda_a$, $10\lambda_a$, and for the doublet presented in Fig.\ref{fig:2}(a).}
    \label{table:2}
    \end{table}

\section{I\MakeLowercase{ntegrated acoustic spiral delay lines with $\approx$ 2.5 \si{\us} on-chip signal delays}}\label{sec3}

\begin{figure}[!hbtp]
{\includegraphics[width = 1.0 \columnwidth]{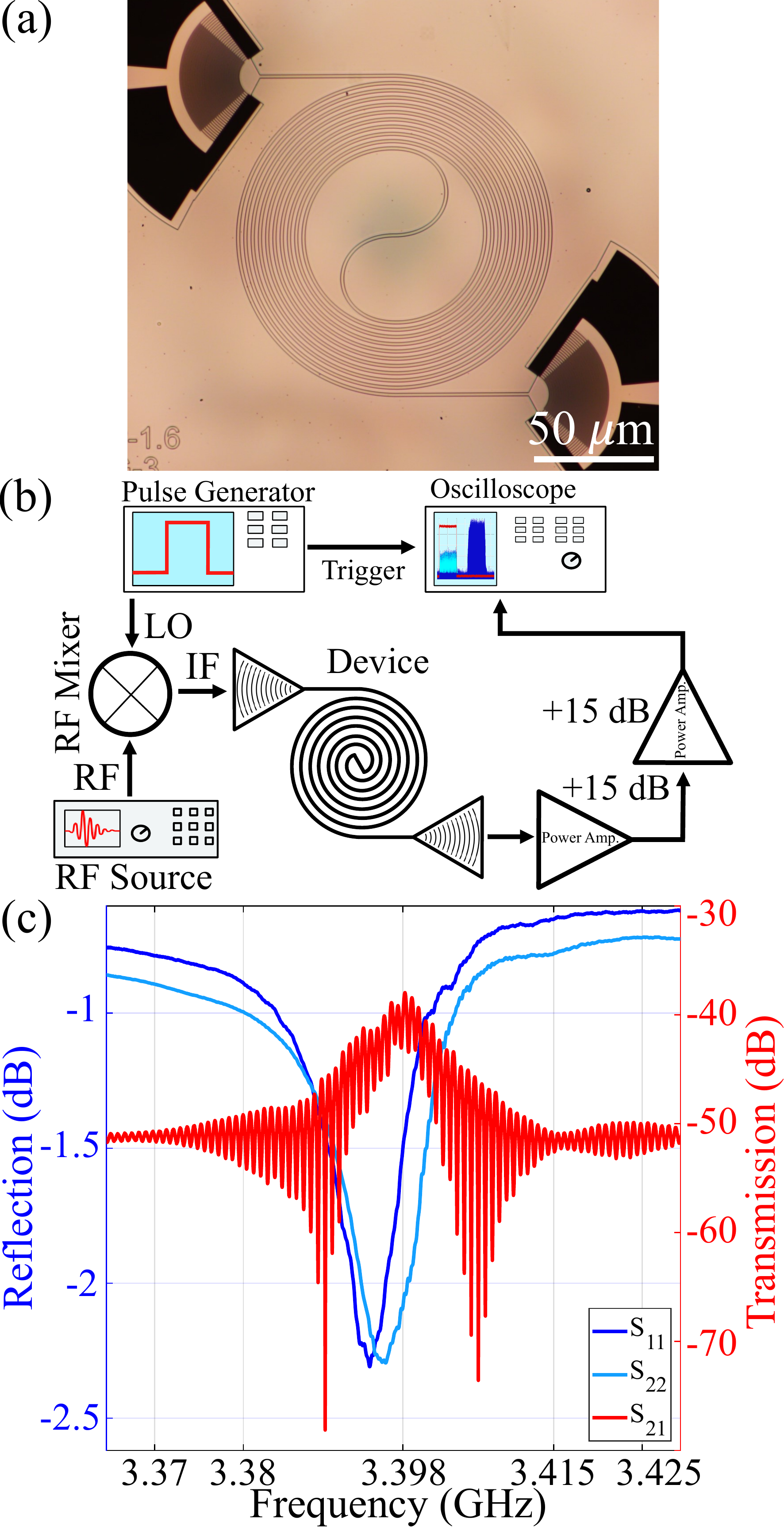}}
  \caption{(a) Optical image of a GaN spiral delay line with a 5.5 \si{\mm} spiral length and 3 \si{\um} gap between successive turns of the spiral. The IDTs both have a $1.6 $ \si{\um} period and an angular span of \qty{90}{\degree}. (b) Schematic of the time-domain measurement setup. The RF signal from a signal generator at the IDT frequency is shaped by mixing with a pulse of a variable period, pulse width, and duty cycle. The shaped signal is sent to the transmitter IDT, propagates through the delay line, and is picked up by the receiver IDT. It is amplified and displayed on a high-speed oscilloscope. The scope is triggered by the pulse generator for synchronization of the time scans. (c) Measured S-parameter spectra through one of the spiral delay lines using a VNA. The transmission ($S_{21}$) and reflection ($S_{11}$ and $S_{22}$) are indicated in red, blue, and light blue colors. The spiral length was 4.1 \si{\mm} and the peak transmission was $\approx$ \qty{-38}{\dB} at a transmission frequency of $\approx$ \qty{3.4}{\GHz}.}
  \label{fig:4}
  \end{figure}

As a demonstrator application for this low-loss PnIC platform, and to show the potential of being able to route GHz sound waves in \si{\um}-scale on-chip waveguides, we build long spiral delay lines that can achieve $>$ \qty{2.5}{\us} RF signal delays in a compact chip-scale footprint. Given that the speed of sound is $\approx10^{-5}$ the speed of light, acoustic delay lines provide a practical route toward achieving large RF delays ($\gg$ 100 ns) in chip-scale platforms. Acoustic delay lines have a long history in signal processing \cite{datta1986surface, campbell2012surface}, and are recently undergoing a resurgence with the development of new piezoelectric platforms like LN on insulator \cite{lu2018s0}, which can support waveguiding using suspension. As outlined above, the GaN platform's main advantage is the prospect of integrating signal processing circuitry on the same die and exerting dynamic control on sound waves in \si{\um}-scale geometries by exploiting acoustoelectric interactions.

\begin{figure}[!hbtp]
{\includegraphics[width = 1.0 \columnwidth]{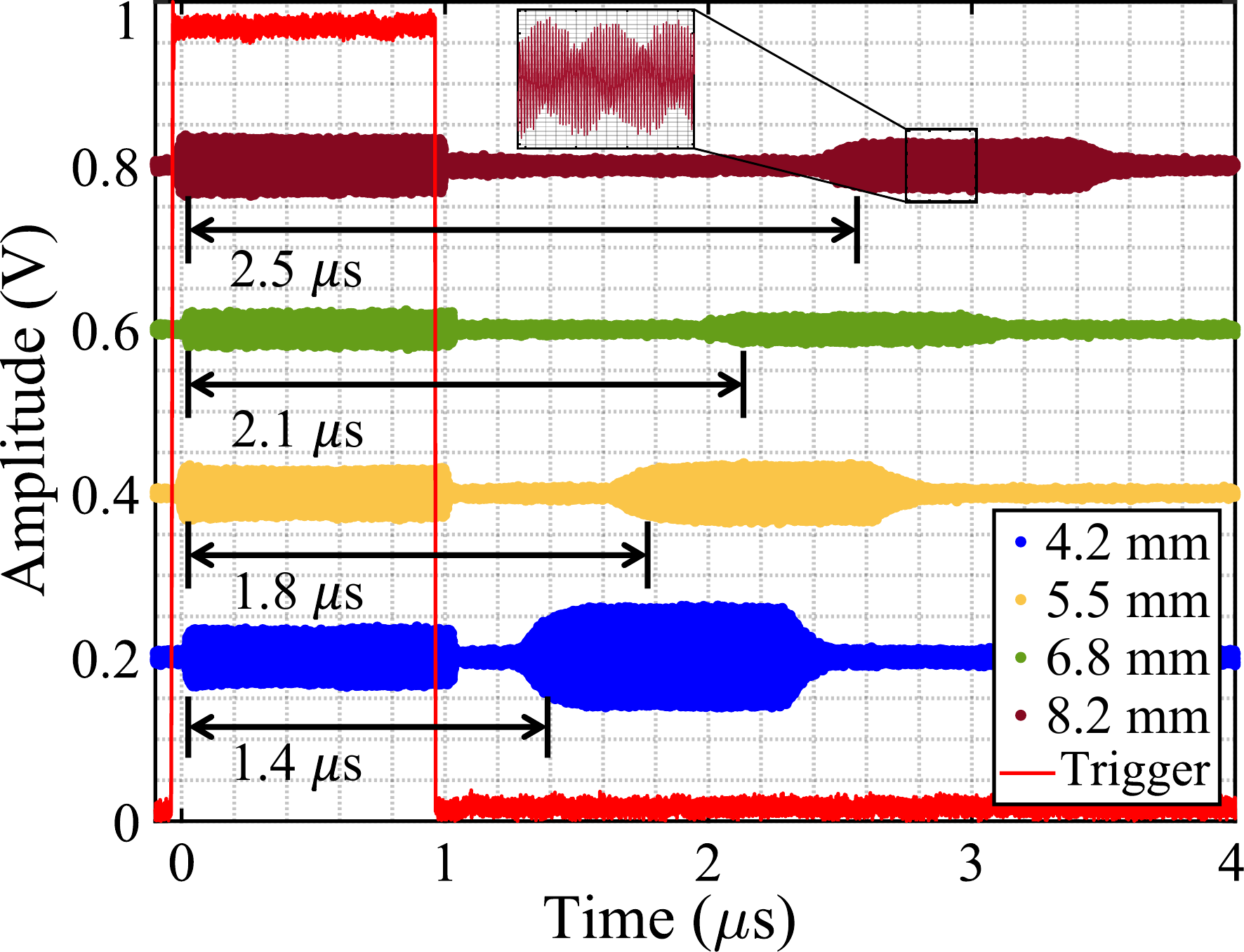}}
  \caption{ RF signal delay measurements for delay lines with lengths of 4.2 \si{\mm} (blue), 5.5 \si{\mm} (yellow), 6.8 \si{\mm} (green), and 8.2 \si{\mm} (brown). The measured signal delays are 1.4 \si{\us},  1.8 \si{\us},  2.1 \si{\us}, and 2.5 \si{\us}, respectively. The input pulse is shown in red for reference.}
  \label{fig:5}
\end{figure}

\begin{figure}[!hbtp]
{\includegraphics[width = 1.0 \columnwidth]{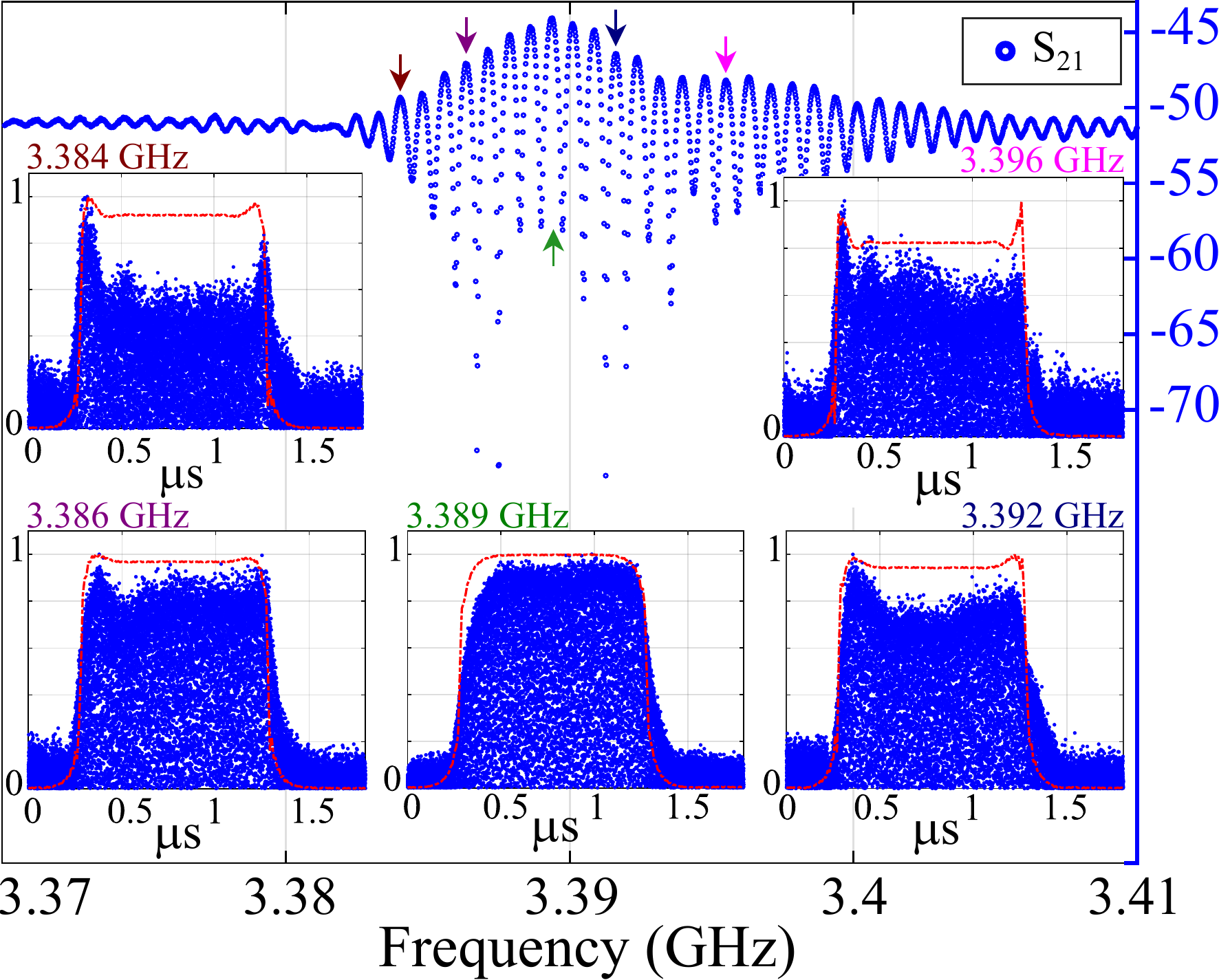}}
  \caption{ Measured transmission ($S_{21}$) spectrum of a spiral delay line with 4.2 \si{\mm} length. The insets show the received acoustic pulse shape as the input RF frequency is varied across the $S_{21}$ spectrum with the frequency locations on the spectra indicated by colored arrows. The predicted pulse shape from linear response theory is overlaid in red, showing reasonable qualitative agreement with the measured results.}
  \label{fig:6}
\end{figure}

Figure \ref{fig:4}(a) shows an optical microscope of a fabricated spiral delay line with a length of \qty{5.5}{\mm}. Focusing IDTs similar to the ones used in the microring resonator experiment described above, are used to launch and detect \qty{3.4}{\GHz} Lamb waves into and out of the spiral waveguide. The measured RF transmission ($S_{21}$) through the spiral waveguide ($L_{sp}$ = \qty{4.1}{\mm}), along with the IDT characteristics ($S_{11}$, $S_{22}$) are shown in Fig.\ref{fig:4}(c). The transmission spectrum is clearly peaked at the IDT $S_{11}$ dips showing acoustic wave transmission through the length of the spiral. We can also verify that the acoustic wave makes its way all the way to the center of the spiral and out (and eliminate leaky modes and direct cross-talk), by fabricating spirals with the central connection missing. Appendix \ref{secA3} shows that the signal transmission $S_{21}$ drops to the vector network analyzer (VNA) noise floor and serves as our control experiment. An interesting feature of the transmission spectrum for the intact spiral in Fig.\ref{fig:4}(c) is the presence of sharp periodic dips (zoomed-in image in Fig.\ref{fig:6}), which are due to an unintended acoustic Fabry-Perot cavity being set up inside the spiral delay line due to reflections. Estimating the free spectral range (${\Delta}f$) from the transmission and using the group velocity of the Lamb waves in the resonator (from the delay measurement Fig.\ref{fig:5}) ($v_g$), we estimate the cavity length ($L_{cav}= v_g/2{\Delta}f$) to be $\approx$ $L_{sp}/2$. 

Figure \ref{fig:4}(b) shows the measurement setup used to perform the time delay measurements using the acoustic spiral delay lines \cite{balram2017acousto}. A continuous wave RF signal from a signal generator (at the IDT frequency) is mixed with a short pulse from a pulse generator. This pulsed signal is sent to the transmitter IDT, where it gets converted to sound and is launched into the spiral waveguide. At the output of the spiral waveguide, a nominally identical IDT converts the acoustic signal into an electrical signal, which is amplified and fed to a high-speed oscilloscope. The scope trace is triggered by the pulse generator to synchronize the scans and allow for averaging to improve the received signal-to-noise ratio. In our experiments, we use a pulse width of  \qty{1}{\us} and a repetition rate of \qty{100}{\kHz}. 

Figure \ref{fig:5} shows the measured delays from four different delay lines with spiral lengths varying from 4.2 \si{\mm} to 8.2 \si{\mm}, with the corresponding signal delays ranging from \qty{1.4}{\us} to \qty{2.5}{\us}. The reference signal from the pulse generator is shown overlaid in red, and the received signal is shown in a different color for each device. The scope trace for the received signals shows two components, a fast signal that is overlapped with the trigger which originates from (weak) electromagnetic radiation cross-talk between the transmitter and receiver IDTs, and a slower acoustic signal that arrives later that corresponds to sound propagating through the spiral waveguide. From the shape of the received signal compared to the input pulse, it is clear that the IDT and the spiral delay line in combination modify the shape of the transmitted pulse.

As Figure \ref{fig:5} shows, the longest spiral delay lines provide an on-chip delay of $\approx$ \qty{2.5}{\us}, which corresponds to a free-space electromagnetic delay of $\approx$ \qty{750}{\m}, or equivalently a coaxial cable delay of $\approx$ \qty{520}{\m}, all within an on-chip footprint $<$ \qty{0.25}{\mm^2}, which shows the real promise of guiding sound in \si{\um}-scale waveguides. From the measured spirals S-parameter peak transmission ($S_{21}$), we can estimate our on-chip propagation loss to be $\approx$ \qty{3.6}{\dB\per\mm}, which agrees reasonably well with the loss values extracted from the microring resonator $Q$ measurements in Table \ref{table:2}.

The combination of the IDT and the spiral delay line has strong frequency dispersion which affects the received pulse shape, as shown in Fig.\ref{fig:5}. The shape of the received pulse is also sensitive to the frequency as shown in Fig.\ref{fig:6} which presents the transmitted ($S_{21}$) signal through a spiral delay line of length \qty{4.2}{\mm}. The regular periodic dips corresponding to the unintended acoustic Fabry Perot cavity formed in the delay line can be clearly seen. To study the dispersive effects of the delay line system on the pulse shape, we perform the signal delay experiment shown in Fig.\ref{fig:4}(b), but now tune the frequency of the signal generator to different locations in the $S_{21}$ spectrum, indicated by the colored arrows in Fig.\ref{fig:6}. The received acoustic pulse shape corresponding to each case is shown in the inset in Fig.\ref{fig:6}. As can be seen, the operating frequency has a significant effect on the received pulse shape. 

To model the system response, we use methods from linear systems response theory \cite{balram2017acousto} to decompose the input signal into its Fourier components, propagate each frequency component through the system and recombine at the output using an inverse Fourier transform to recover the time domain dynamics.  We add an additional frequency-dependent loss term which allows us to replicate the observed qualitative time domain behavior, see Appendix \ref{secA2} for further details. The frequency dependence of the loss term broadly follows the IDT spectral response, which indicates the relative efficiency with which the different frequency components are launched as propagating sound waves into the spiral waveguide. The strongly dispersive nature of these spiral delay lines and more generally, the strong geometric dispersion accessible in these \si{\um}-scale waveguide platforms provide new avenues for spectrum shaping and phase matching to harness acoustic nonlinear interactions like four-wave mixing \cite{mayor2021gigahertz}.

\section{C\MakeLowercase{onclusions}}\label{sec5}

We have shown that GaN-on-SiC supports low loss GHz frequency acoustics in \si{\um}-scale waveguides and resonators with resonator loss measurements exceeding the isotropic material Akhiezer damping limit and long spiral delay lines with on-chip RF signal delays exceeding \qty{2.5}{\us}. Given GaN's established technology base as a high electron mobility semiconductor, our work opens up the prospect of engineering traveling wave acoustoelectric interactions in waveguide and resonator geometries with associated implications for on-chip RF signal processing, in particular the implementation of novel wireless protocols like full-duplex wireless.

\section{A\MakeLowercase{cknowledgments}}

We would like to thank Stefano Valle, Jacob Brown, Fahad Malik, Martin Cryan, Martin Kuball, Bruce Drinkwater, John Haine, and Hugues Marchand for their valuable discussions and suggestions. The authors acknowledge funding support from the European Research Council (ERC-StG SBS 3-5, 758843) and the UK's Engineering and Physical Sciences Research Council (GLIMMER, EP/V005286/1). Nanofabrication work was carried out in the Bristol cleanroom using equipment funded by an EPSRC quantum technology capital equipment grant (QuPIC, EP/N015126/1).  

\begin{appendix}

\section{Temporal Coupled-Mode theory for analysis of acoustic microring resonators} \label{secA1}

\par
\begin{figure}[!]
\centering
{\includegraphics[width = 1 \linewidth]{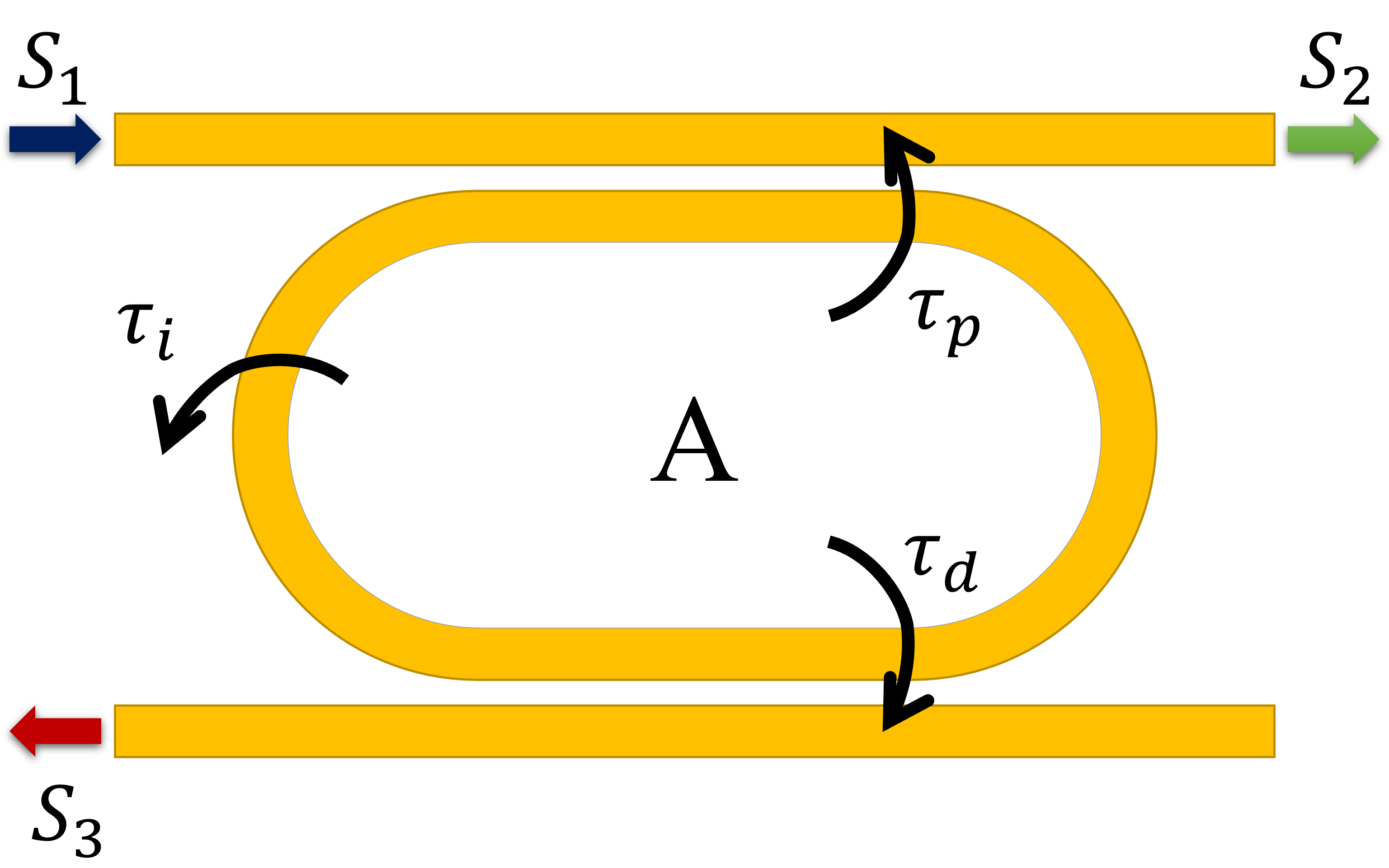}}
  \caption{ An abstract diagram showing the critical components of the 4-port system. Single mode input into system $S_1$, through output $S_2$, and drop output $S_3$. While field amplitude in the microring is $A$ and $\tau_d$, $\tau_p$, and $\tau_i$ are the coupled decay rates from the cavity to through-port, drop-port, and the microring intrinsic loss, respectively.}
  \label{fig:S1}
\end{figure}

 To quantify the waveguide resonator interaction in analogy with integrated photonics, we employ temporal coupled mode theory following the notation of \cite{jd2008photonic}. A schematic of the add-drop waveguide coupled racetrack resonator is shown in Fig.\ref{fig:S1} denoting the various field amplitudes involved in the coupled mode equations described below.

\par Assuming a circulating field amplitude $A$ in the resonator and $S_{i}$ in the $i$-th waveguide port, we can formulate the standard coupled mode equations following \cite{jd2008photonic}:

\begin{equation} \label{eqn9}
    \frac{dA}{dt} = -i\omega_0 A - \frac{A}{\tau_{d}} - \frac{A}{\tau_{p}} - \frac{A}{\tau_{i}} +  \sqrt{\frac{2}{\tau_{p}}}S_1 
\end{equation}

\begin{equation}
    S_2 = S_1 + \sqrt{\frac{2}{\tau_{p}}}A
\end{equation}

\begin{equation}
    S_3 = \sqrt{\frac{2}{\tau_{d}}}A
\end{equation}\\

where the $\tau_{i}$ represent the various decay channel time constants (intrinsic and coupling to top and bottom waveguides) and are related to the respective quality factors by $Q_{i}=\omega_{0}\tau_{i}/2$.

For harmonic excitation at a frequency $\omega$, we can solve these equations to extract the drop port (power) transmission $S_{31}$ as :

\begin{equation}
    T_{31}(\omega) = \frac{|S_3|^2}{|S_1|^2} =  \frac{\frac{4}{\tau_p \tau_d}}{(\omega-\omega_0)^2 + (\frac{1}{\tau_{d}} + \frac{1}{\tau_{p}} + \frac{1}{\tau_{i}})}
\end{equation}\\

\par For simplification, we assume identical top and bottom waveguide decay rates $\tau_d = \tau_p = \tau_{w}/2$, where $\tau_w$ is the net waveguide decay lifetime. The total decay rate (expressed in lifetimes) and corresponding total $Q$ are then:

\begin{equation}
    \frac{1}{\tau_{cav}} = \frac{1}{\tau_w}+ \frac{1}{\tau_i}
\end{equation}

\begin{equation}
    \frac{1}{Q} = \frac{1}{Q_w}+ \frac{1}{Q_i}
\end{equation}\\
In terms of these quantities, the drop-port transmission spectrum becomes:

\begin{equation}
    T_{31}(\omega) = \frac{\frac{1}{4Q_w^2}}{\left(\frac{\omega-\omega_0}{\omega_0}\right)^2 + \frac{1}{4Q^2}}
\end{equation}\\

\par Now, we can analyze the $Q_i$ and $Q_w$ through the measured total $Q$ and the peak transmission spectrum ($T_{31}(\omega_0)$):

\begin{equation}\label{eqn10}
    T_{31}(\omega_0) = \left ( \frac{Q}{Q_w} \right)^2 =  \left(\frac{Q_i}{Q_w+Q_i}\right)^2
\end{equation}\\

\par which leads us to

\begin{equation}
    Q_w = {\frac{Q}{\sqrt{T_{31}(\omega_0)}}}
\end{equation}
\begin{equation}
    Q_i = \frac{QQ_w}{Q_w - Q}
\end{equation}

\section{Linear Response Theory for modeling spiral delay line transmission and dispersion} \label{secA2}

\begin{figure}[!htbp]
\centering
{\includegraphics[width = 1 \columnwidth]{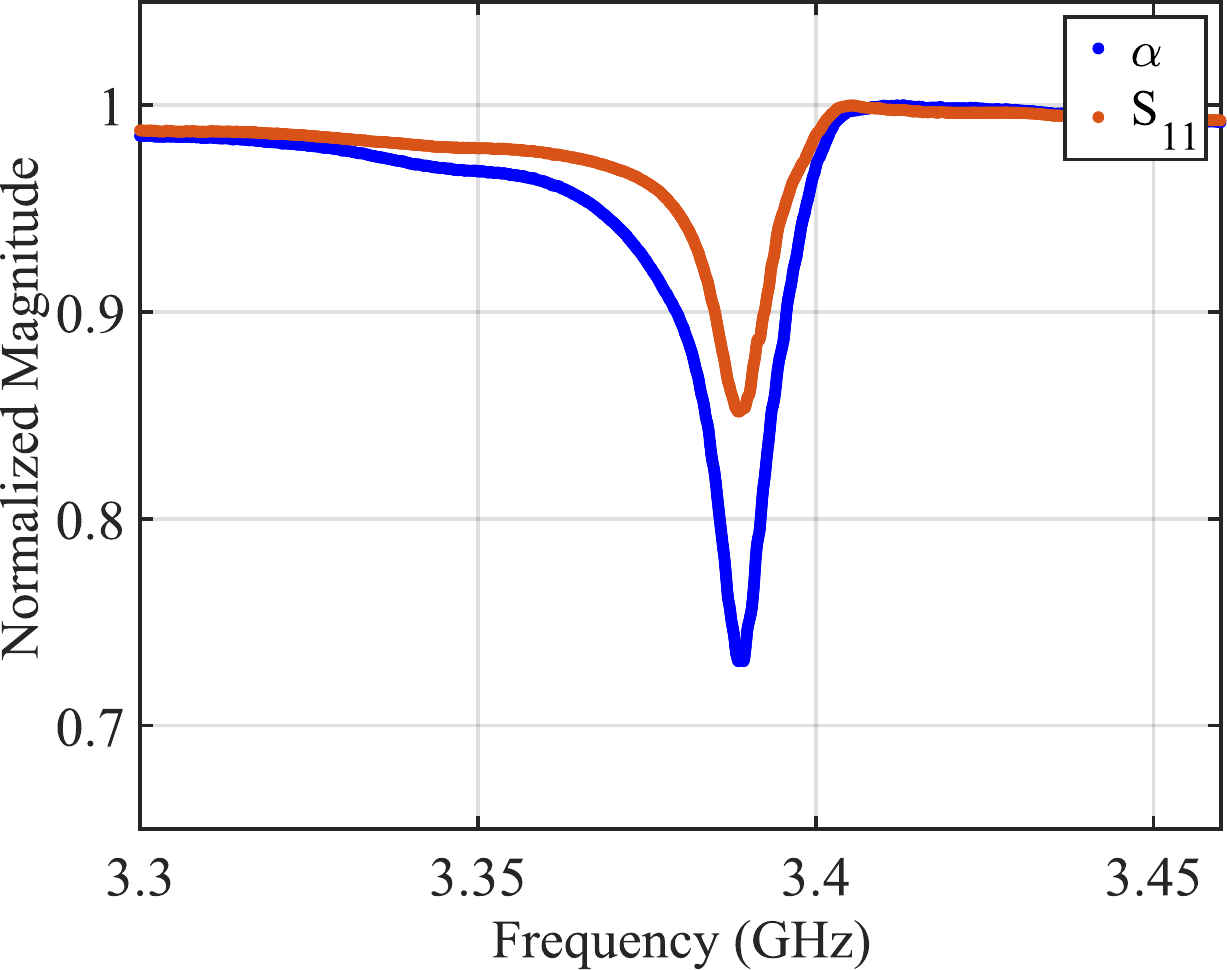}}
  \caption{ Phenomenological $\alpha$ introduced into the model to fit the observed time domain pulse shapes.}
  \label{fig:S2}
\end{figure}

\par To model the pulse shape of the received time domain signal that traverses the spiral delay line, we follow a standard procedure from linear systems analysis by decomposing the input signal into its constituent frequency components by a Fourier transform, propagating each component through the waveguide and recombining the signal at the received IDT by an inverse Fourier transform. The input is a pulsed RF signal $(H_i(t))$ of 1 \si{\us} pulse width and \qty{3.4}{\GHz} center frequency. The IDT is modeled as a frequency-dependent acoustic wave transducer that acts on each frequency component ($h_{i}(\omega)$) of the input signal according to its measured complex amplitude (voltage) reflection spectrum $S_{11}$, generating a corresponding acoustic wave amplitude ($u_i(\omega)$):

\begin{figure}[!htbp]
\centering
{\includegraphics[width = 1 \linewidth]{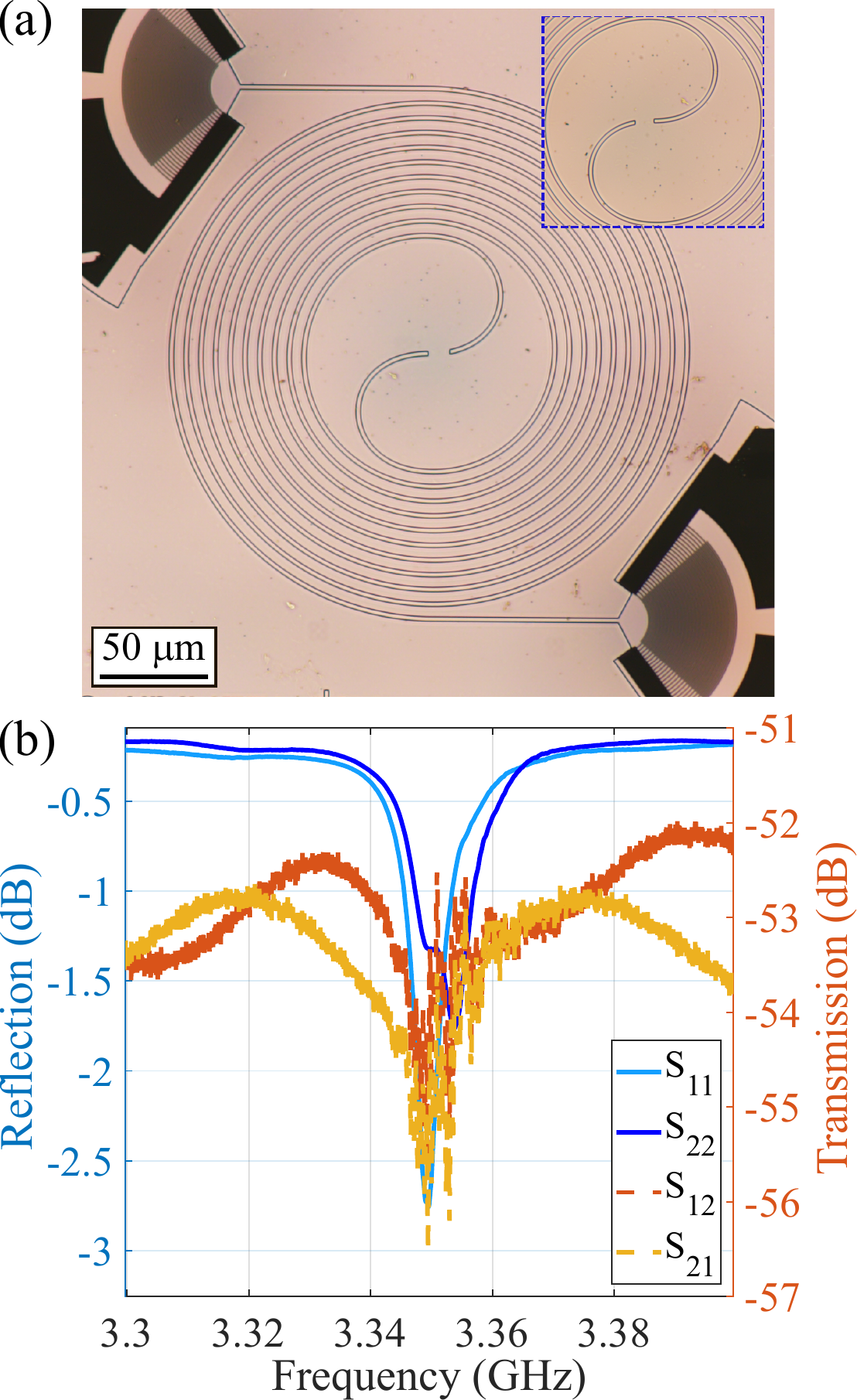}}
  \caption{ The disconnected spiral delay line, (a) An optical image of the fabricated device, and the inset is the close look at the disconnected center of the spiral waveguide, (b) The measured S- parameters of the device, $S_{21}$ and $S_{12}$ are the transmission while $S_{11}$ and $S_{22}$ are the responses of transmitter and receiver IDTs.}
  \label{fig:S3}
\end{figure}

\begin{equation}
    u_i = h_i(\omega)(1-S_{11}(\omega))
\end{equation}\\
The acoustic wave is propagated through the spiral waveguide and the output wave amplitude $(u_o)$ can be represented as:

\begin{equation}
    u_o(\omega) = u_i(\omega)e^{i(\beta(\omega)+i\alpha(\omega))L_{sp}}
\end{equation}\\

where $\beta$ is the wave propagation constant, $L_{sp}$ is the spiral waveguide length, and $\alpha$ is the propagation loss coefficient per unit length. To reproduce the measured pulse shapes, our $\alpha(\omega)$ is not a constant, but has the frequency variation as shown in Fig.\ref{fig:S2}. The excess loss and the close mirroring of the IDT lineshape indicate that the acoustic power launched by the IDT off-resonance is not propagating, although this needs to be verified further.

At the receiving IDT, the signal amplitude $u_o(\omega)$ is reconverted back into the electrical domain ($h_o(\omega)$) using the receiving IDT spectrum ($S_{22}(\omega)$. Finally, the received signal ($h_{o}(\omega)$) is converted back into the time domain ($H_o(t)$) by an inverse Fourier transform. The red curves shown in Fig.\ref{fig:6} in the main text are evaluated in this way. We would like to note here that the agreement between theory and experiment here is qualitative. While we are able to reproduce the received pulse shapes, the predicted signal amplitudes (red curves in Fig.\ref{fig:6}) have been scaled to fit the experimental data.

\section{Disconnected Spiral Delay Line} \label{secA3}

\par To eliminate acoustic cross talk and leaky modes in  the long spiral waveguide and to confirm the observation that the acoustic wave makes its way all the way to the center of the spiral and out, we fabricated a spiral delay line with a disconnected center. The broken center device is shown in Fig.\ref{fig:S3}(a) and there is a 10 \si{\um} gap between two ends of the waveguide (inset of Fig.\ref{fig:S3}(a)). The spiral waveguide is 5.9 \si{\mm} long and IDTs have a 1.6 \si{\um} period. Figure \ref{fig:S3}(b)  shows the measured IDT responses ( $S_{11}$ and $S_{22}$) which behave nominally as expected. The measured transmission spectra are however  below -50 dB which is almost at the VNA noise floor (Fig.\ref{fig:S3}(b)) and should be compared to the traces in Fig.\ref{fig:4}(c).

\end{appendix}

\bibliography{References}

\end{document}